%% file: mods_multiple_v5_resub_arxiv.tex
\documentclass[aps,prb,reprint]{revtex4-1}
\usepackage[usenames,dvipsnames]{color}
\usepackage{amsmath,array,amssymb}
\usepackage{graphicx}
\usepackage{hyperref}
\usepackage{textcomp}
\usepackage{braket}
\usepackage{multirow}
\usepackage{indentfirst}
\usepackage[FIGTOPCAP,raggedright,nooneline,bf,footnotesize]{subfigure}
\usepackage{paralist}
\usepackage{overpic}
\usepackage{soul} 
\usepackage{xspace} 

\usepackage{comment}
\usepackage{tablefootnote}
\usepackage{colortbl}
\definecolor{LightCyan}{rgb}{0.88,1,1}
\definecolor{Plum}{rgb}{.95,.7,.95}

\newcommand{\cmt}[1]{}

\renewcommand{\vec}[1]{\boldsymbol{#1}}

\renewcommand{\eqref}[1]{Eq.~(\ref{#1})}

\newcommand{\eryso}[0]{{Er$^{3+}$:Y$_2$SiO$_5$}\xspace}

\newcommand{\yso}[0]{Y$_2$SiO$_5$\xspace}
\newcommand{\er}{Er$^{3+}$\xspace}
\newcommand{\nd}{Nd$^{3+}$\xspace}
\newcommand{\yb}{Yb$^{3+}$\xspace}
\newcommand{\ce}{Ce$^{3+}$\xspace}
\newcommand{\y}{Y$^{3+}$\xspace}
\newcommand{\yvo}[0]{YVO$_4$\xspace}
\newcommand{\ylf}[0]{YLiF$_4$\xspace}
\newcommand{\cawo}[0]{CaWO$_4$\xspace}

\graphicspath{{img/}}

\begin{document}


\title{Superhyperfine induced photon-echo collapse of erbium in \yso}

\author{B. Car$^1$}
\author{J.-L. Le Gou\"et$^1$}
\author{T. Chaneli\`ere$^2$}
\affiliation{$^1$Laboratoire Aim\'e Cotton, CNRS, Univ. Paris-Sud, ENS-Cachan, Universit\'e Paris-Saclay, 91405, Orsay, France}
\affiliation{$^2$Univ. Grenoble Alpes, CNRS, Grenoble INP, Institut N\'eel, 38000 Grenoble, France}

\date{\today}

\begin{abstract}
We investigate the decoherence of \er in \yso at low magnetic fields using the photon-echo technique. We reproduce accurately a variety of the decay curves with a unique coherence time by considering the so-called superhyperfine modulation induced by a large number of neighbouring spins. There is no need to invoke any characteristic time of the spin fluctuations to reproduce very different decay curves. The number of involved nuclei increases when the magnetic is lowered. The experiment is compared with a model associating 100 surrounding ions with their exact positions in the crystal frame. We also derive an approximate spherical model (angular averaging) to interpret the main feature the observed decay curves close to zero-field.

\end{abstract}


\maketitle

\section{Introduction}

The interest for rare-earth doped materials has been recently renewed by the quest for quantum technology devices. The longest coherence times are generally observed in non-Kramers ions (even number of electrons) because their spins possess a nuclear spin character in low symmetry site as the emblematic europium in \yso \cite{zhong2015optically}. The research activity on Kramers ions has been maintained, despite the generally lower coherence times, because they cover an interesting wavelengths panel in the infrared region. Erbium holds a lot of promises in that sense because of the compatibility with the optical fiber communication range. The large electron spin undeniably induces decoherence but offers also significant advantages that have been reconsidered for quantum information processing. The electron spin resonance (ESR) falls in the GHz range, an actively investigated region to operate superconducting qubits, allowing an hybridization between quantum circuits and spin ensembles \cite{Kubo}. The apparently detrimental spin sensitivity appears as a major benefit when a very low number of spins is targeted \cite{Probst}. The interplay between optics and microwave is fully exploited by the scheme of coherent frequency conversion. Indeed, the transduction of quantum states from the microwave to the optical domain appears as a missing link in the quantum technology landscape \cite{Lambert2020, Lauk_2020}. The potential of Kramers ion has been rapidly identified in this context \cite{Fernandez_Gonzalvo}.

We focus on the low magnetic field region, typically below 100mT. This offers fundamental interests beyond the experimental advantage of using smaller magnets. First, the phonon density is small at cryogenic temperature (2-4K). Second, the ESR transitions (potentially involving the hyperfine structure) fall in the few GHz range and are then directly compatible with superconducting high-Q resonators \cite{Lambert2020, Lauk_2020}. Early demonstrations have already involved \eryso \cite{PhysRevLett.110.157001}. Optical spin excitation can also be obtained conveniently with a single laser modulated with modern electrooptics devices. Many experiments are performed  close to zero-field in practice with \er or \yb for example \cite{Chen, PhysRevB.98.195110, PhysRevLett.124.053606, PhysRevB.97.064409}.

Despite a clear interest for low magnetic field region, the variation of the coherence time is essentially unexplored. One has the tendency to dodge the issue by noting that echos exhibit strong  superhyperfine modulations induced by neighbouring ligands nuclei (sometimes called ligands interaction) making a complete analysis difficult because of the diversity of modulation patterns. In any case, the erratic nature of the measurements disappears close to zero-field where modulations are absent and the decay (exponential or not) is extremely rapid although one does not expect any change in the spin bath dynamics. This coherence collapse may give the impression that a sudden change of regime is happening. This is not the case. We will give a unified vision of the decoherence at low field without invoking any change in the spins dynamics and precisely explain the collapse of the measured coherence time when field is reduced. Our analysis is based on the superhyperfine coupling exclusively \footnote{We will keep the term superhyperfine to designate the electron spin to ligand interaction even in absence of hyperfine structure as for the even isotopes of \er}. Again, we do not consider the ligand nuclear flip-flops nor the electron spin flip-flops that induce a magnetic noise and affect the coherence time of the impurity. We investigate a different mechanism, static in the sense that we neglect the spin dynamics.

We consider a large collection of yttrium surrounding an \er center. They all have different couplings to the dopant electron spin because of the distance and the anisotropy of the \er dipolar field. The sudden excitation by the brief echo measurement pulses of those multiple frequencies leads to an {\it apparent} decay time that is much shorter that the coherence time induced by the background spin flip-flops observed at a larger field \cite{zhong2015optically, liu2006spectroscopic, mims1972electron}. This phenomenon has been discussed early for ESR transitions \cite{hurrell1971nuclear, mims1972electron}. It explains the shortness, literally the collapse of the measured dephasing times for the different Kramers ion (including \er) at zero-field \cite{liu2006spectroscopic}.

Although early predicted in ESR \cite{hurrell1971nuclear, mims1972electron} as an extreme case of envelope modulation, the superhyperfine nuclear induced decay regime is rarely observed in practice, because the experiment are usually performed in the radio-frequency X-band ($\sim$ 9GHz). In that case, the magnetic field is already sufficiently large ($>$100mT) to dominate the electron spin dipolar field \cite{guillot-noel_direct_2007}, so the superhyperfine modulations are weakly contrasted (but visible in the Fourier spectrum). Custom-made ESR spectrometers with a variable resonant frequency are clearly more adapted \cite{probst2020hyperfine}. Optical techniques are intrinsically broadband and can be implemented close to zero-field without modification of the laser set-up \cite{mitsunaga_cw_1990,car2017selective}. Using the photon-echo technique, we show that the rapid decoherence of \eryso is well explained by the static superhyperfine interaction with a collection of yttrium ions.

Historically, the superhyperfine interaction between a Kramers ion and the ligand nuclei has been widely studied starting from the seminal work of Mims on \ce in \cawo \cite{PhysRev.137.A61}. It has been also evidenced later-on in \eryso using standard ESR techniques \cite{guillot-noel_direct_2007} or superconducting resonators \cite{PhysRevB.92.014421}. Optical and RF measurements has allowed to characterize the superhyperfine interaction in a variety of host crystals as \yvo with \yb and \nd \cite{Huan:19_YVO4,PhysRevB.77.125111_YVO4}, \ylf with \nd and \er \cite{PhysRevB.58.5692_YLF, Kukharchyk_2018_YLF} or \cawo with \er exhibiting a remarkably high sensitivity \cite{mr-2020-10}.

Concerning \eryso because of the perspectives in classical and quantum processing, advanced spectroscopic studies have been used to accurately describe the dopant in the crystal-field \cite{li1992spectroscopic, PhysRevLett.123.057401}, the different $g$-tensors (in both substitution sites of yttrium and in the ground and optically excited state of \er) \cite{sun_magnetic_2008}. The hyperfine tensors for odd-isotopes are also known \cite{PhysRevB.74.214409, PhysRevB.97.024419}. This abundant literature is essential to describe the superhyperfine coupling.
In this study, we crudely extract the \y positions from the crystal structure \cite{YSOstructure} and calculate the couplings one by one generalizing our previous approach in \cite{car2017selective}.

The paper is organized as follows. We first describe the experimental apparatus and show typical photon-echo decay curves between 0 and 133 mT. We analyse the curves by extending the envelope modulation model to a cluster of 100 surroundings yttrium ions. We show that for a decreasing magnetic field, the number of coupled nuclei increases. We finally interpret the low-field collapse by assuming an equivalent homogeneous spherical distribution of yttrium around the rare-earth ion. This allows to reproduce the main feature of the observed decay curve and to introduce the notion of an inflating sphere of influence of the \er ion when the magnetic field is reduced.


\section{Experiment}
We perform 2-pulse photon echo measurements on the $^{4}I_{15/2} \rightarrow ^{4}I_{13/2}$  transition of \er in \yso. The configuration is very similar to our previous study of the superhyperfine coupling with a single nucleus \cite{car2017selective}. As a reference frame for the magnetic field orientation, we use the optical frame $(D_1,D_2,b)$ \cite{YSOstructure, li1992spectroscopic} where $D_1$ and $D_2$ are the extinction axes. This is a natural frame for optical measurements. Additionally, the axes are perpendicular as opposed to the monoclinic crystal frame.

We roughly orientate the magnetic field $\vec{B}$ in the $(D_1,D_2)$ plane at $50^{\circ}$ from $D_1$ by rotating the crystal in the magnet close to the previously studied configuration \cite{car2017selective}. Staying in the $(D_1,D_2)$ plane simplifies the analysis because the so-called magnetic subsites (related by a $C_2$ symmetry about ${b}$) are equivalent. This aspect will be discussed in the appendix \ref{appendix:C2}.

\begin{figure}[h]
\centering
\includegraphics[width=.8\columnwidth]{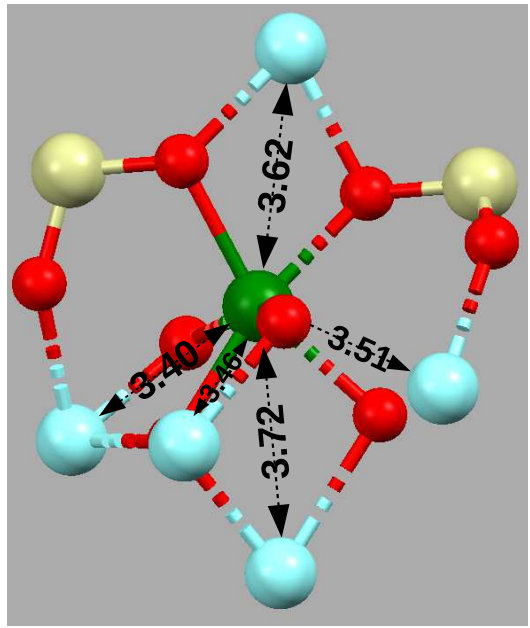} 
\caption{Reduced crystal cell representing the site 1 in the \yso matrix: erbium in green substitutes an yttrium in the 6-coordinate site. Colour code: cyan - yttrium, red - oxygen and yellow - silicon. The yttrium ions are the 5 nearest neighbors with distances 3.40\AA, 3.46\AA, 3.51\AA, 3.62\AA\, and 3.72\AA\, from the erbium center. The distances of the two represented silicons are 3.26\AA\, and 3.51\AA.}
\label{fig:YSO_site1_red}
\end{figure}

We optically resolve the lowest to lowest spin state transition of site 1 (at 1536.38~nm, see Fig.\ref{fig:YSO_site1_red} for a cell representation) allowing a well defined orientation of the \er magnetic moment in both ground and excited states. We will also take data at zero-field (by zeroing the magnet current) where this assumption fails. This aspect will be discussed in \ref{sec:interpretation}. The sample is lightly doped  (10 ppm, grown by Scientific Materials Corporation)  to avoid the so-called erbium spin flip-flops that may perturb the echo decay curve  (in the regime of small magnetic fields \cite{bottger_optical_2006}).
We cool down the crystal to $1.8$~K. The light propagates along the $b$-axis of the crystal and the polarization is parallel to $D_2$ to maximise the absorption and the photon-echo signal.

In the following, we neglect the response of the $^{167}$Er isotope (22\% of the dopant concentration, with a nuclear spin of $7/2$) that is broadly spread over a large amount of possible hyperfine transitions \cite{PhysRevB.74.214409} and therefore can be neglected because of the optical selection \cite{car2017selective}. Concerning the nuclear spins present in the matrix, $^{89}$Y is the most abundant (100\% natural abundance with a 2.1 MHz/T nuclear magnetic moment).  $^{29}$Si is also present with 4.7\% abundance and a 8.5 MHz/T magnetic moment \footnote{$^{17}$O with non-zero nuclear moment appears only as traces with 0.04\% abundance.}
It is important to keep in mind that the natural abundance scales the modulation contrast which cannot be larger than 4.7\% for $^{29}$Si \cite{kevan1979time}. In any case, the $^{29}$Si nuclear modulations are typically $\sim$20 times weaker than the $^{89}$Y. Additionally, because of a larger magnetic moment, $^{29}$Si nuclear modulations would appear at a four times larger frequency (as the ratio of nuclear moments) making them difficult to observe at our measurement time scale. That is the reason why we focus on yttrium nuclei exclusively in the following.

\begin{figure}[h]
\centering
\includegraphics[width=\columnwidth]{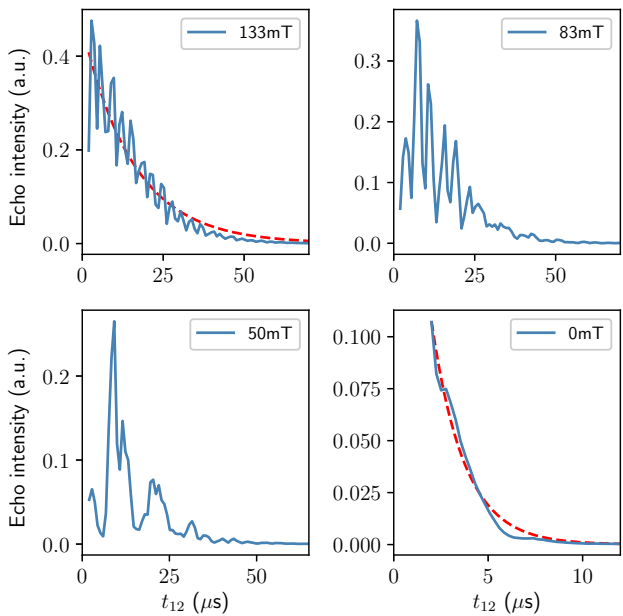} 
\caption{Echo decay curves for $B=133$, $83$, $50$ and $0$ mT exhibiting different modulation patterns. We fit data by an exponential decay (red dashed line) for $B=133$ and $0$ mT and find decay times of $63\, \mu s$ and  $6.9\, \mu s$ respectively. The meaning of theses characteristic times will be discussed in details.}
\label{fig:mods_fig_exp}
\end{figure}

We plot the decay curve of the photon-echo signal as a function of $t_{12}$, the delay between the 2 pulses, for different values of the magnetic field (see Fig.\ref{fig:mods_fig_exp}).  It is important to note that we cannot obtain the echo intensity for $t_{12}<2 \mu$s because the signal is overwhelmed by the strong pulse free-induction-decay. At $B=133$ mT, the envelope ($\sim 63\, \mu s$ decay time) is clearly modulated. Below this value, the pattern is erratic. At $B=0$ mT, we retrieve a smoother but faster decay curve which could be interpreted as an exponential decay at first sight with a characteristic time of $6.9\, \mu s$. Theses characteristics have been already reported in the literature even if the intermediate region (between 0 and 100 mT) is usually not considered because of the erratic aspect of the curves and the difficulty to model the envelope modulation. We will tackle this problem and fit data by assuming a single value of coherence time, common to the different curves (from 0 and 133 mT in our case).

%

\section{Model}\label{sec:model}

As discussed in the introduction, we won't consider the spin dynamics that induce decoherence on a longer time-scale that the one we observe at larger fields. This $T_2$ is an upper limit of our echo decay time. As we will discuss later, there is no reason for the $T_2$ to vary within our measurement range. On a shorter time-scale, the echo decay is driven solely by the superhyperfine coupling with a multiplicity of \y ions. To reproduce the experimental photon-echo curves, we need to account for a collection of surrounding yttrium ions. The difficulty in \yso comes from the low symmetry of the host matrix. Indeed when the \er - \y interactions are considered, the different positions of the surrounding yttrium ions do not exhibit a symmetrical structure that would simplify the analysis.
Nevertheless, since the \y positions is known from the crystal structure \cite{YSOstructure}, one can add their contributions using the historical ESR formula \cite{Mims_SHF1, kevan1979time}. This is sufficient to obtain a very satisfying theoretical agreement. This approach has been also successfully to describe ESR in glassy materials, with a profusion of disordered sites, which exhibit less contrasted modulations or even rapid decays induced by the multiple modulation frequencies which can still be used to extract a characteristic coupling \cite{glassy_mims, kevan1979time}. This work is a substantial basis for our analysis.
The change of the electron spin dipole moment orientation between ground and excited state of the \er center modifies the magnetic field seen by a given \y ions. The modulation comes from the electron-nuclear spin mixing excited during the echo sequence. For multiple coupled nuclei, the envelope modulation is obtained as a product of single superhyperfine modulations:

\begin{equation}\label{Vtot}
\mathcal{V}_\mathrm{tot}=\prod_i \mathcal{V}_{i}
\end{equation}

where $\mathcal{V}_{i}$ is the modulation due to the yttrium numbered $i$ which reads as

\begin{equation}\label{Vi}
\mathcal{V}_{i}\left(t_{12}\right) = 1-\frac{\rho_{i}}{2} \left[1-\cos\left(\Delta_{i} \, t_{12}\right)\right]\left[1-\cos\left(\Delta^\prime_{i} \, t_{12}\right)\right]
\end{equation}

where $\rho_{i}$ is the branching contrast (using the terminology developed in \cite{car2017selective}), $\Delta_{i}$ and $\Delta^\prime_{i}$ the superhyperfine splittings in the ground and excited states of erbium respectively (lowest spin states of $^{4}I_{15/2} $ and $^{4}I_{13/2}$).  This gives the modulation of the echo field. As we measure the intensity (as opposed to ESR), the decay should be proportional to 

\begin{equation}\label{Modulation_formule}
\left[ \mathcal{V}_\mathrm{tot} \left( t_{12}\right) \right]^2\times \exp\left({-\frac{4 t_{12}}{T_2}}\right).
\end{equation}

including an exponential decoherence decay and the superhyperfine modulations. A calculation of the parameters $\rho_{i}$, $\Delta_{i}$ and $\Delta^\prime_{i}$ in \eqref{Vi} has been detailed in ref.\cite{car2017selective} for a given yttrium position, a given orientation and magnitude of the magnetic field. This calculation will be briefly summarized as a reminder in appendix \ref{appendix:C2}. We consider a cluster of 100 nearest yttriums positioned in the crystal frame \cite{YSOstructure} and repeat the calculation for each ion $i$ to evaluate $\mathcal{V}_\mathrm{tot}$ (\eqref{Vtot}). Each experimental curve can then be fitted by \eqref{Modulation_formule}. Per curve, there are only two fitting parameters: a vertical scaling factor (normalization) and the $T_2$ value. The term $\left[ \mathcal{V}_\mathrm{tot} \left( t_{12}\right) \right]$ is completely fixed by the magnetic field and the tabulated yttrium positions.

One can even further constraint the fitting parameters by keeping the same value the coherence time $T_2$ within our measurement range 0-133mT. The $T_2$ is induced by the background spin flip-flops (electronic or nuclear depending on the dopant concentration) that should not vary much in our case. For our magnetic field orientation, the maximum electron spin Zeeman splitting is $\sim 9$GHz in the ground state at 133 mT (with a $g$-factor of 4.8 for this orientation), still much smaller than the temperature (1.8K $\sim$ 36 GHz). In absence of net spin polarization, the \er flip-flop rate should not change. The nuclear spins are even less affected by such a weak magnetic field. So the spin dynamics as a whole is essentially unchanged in our range and as a consequence, the $T_2$ should be constant. So we keep the $T_2$ as a free fitting parameter but constrained to be the same for all the curves in the measurement range. We simply introduce a factor of normalization (scaling) between each theoretical and experimental curves.

\begin{figure}[h]
\centering
\includegraphics[width=\columnwidth]{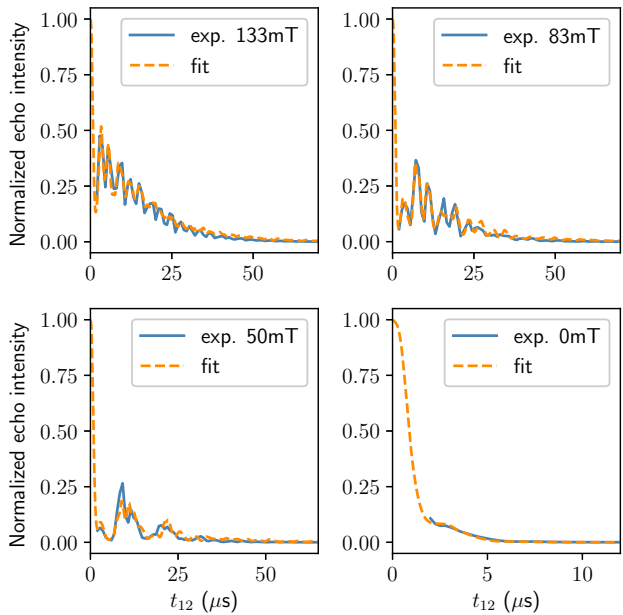} 
\caption{Echo decay curves of Fig. \ref{fig:mods_fig_exp} (blue line) superimposed with the theoretical formula \eqref{Modulation_formule} (orange dashed line). The coherence time is the same $T_2=58 \mu$s.}
\label{fig:mods_fig_fit}
\end{figure}

The match is very satisfactory (see Fig.\ref{fig:mods_fig_fit} and the complementary measurements in \ref{sec:complementary}). We reproduce well the different modulation patterns. The theoretical curves serve as normalization and equals 1 at $t_{12}=0$. All the fitting curves share the same coherence time value $T_2=58 \mu$s despite the disparity of decay patterns. Because we cannot measure the echo intensity for $t_{12}<2 \mu$s, we miss a very rapid decay (at the $\mu$s timescale) that is well-predicted by \eqref{Modulation_formule}. At zero-field, despite the precautions that should be taken when the magnet current goes to zero (see \ref{sec:interpretation}), what we interpret as an {\it apparent} $6.9\, \mu s$ decay in Fig. \ref{fig:mods_fig_exp} seems to be the result of the superhyperfine modulations acting on a much longer $T_2$ decay. We interpret this discrepancy at low-field between the decoherence and the {\it apparent} times as the interaction with the increasing number of coupled nuclei.


\section{Qualitative interpretation}\label{sec:interpretation}

\begin{figure}[h]
\centering
\includegraphics[width=\columnwidth]{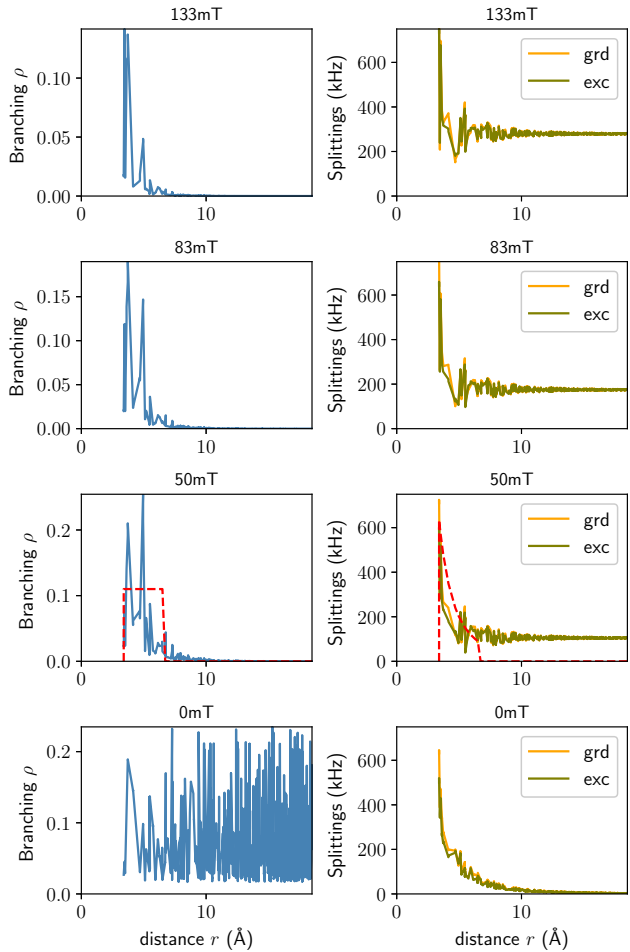} 
\caption{ We plot the modulations parameters $\rho_i$ (left), $\Delta_{i}$ (right, orange) and $\Delta^\prime_{i}$ (right, green) as a function of the nucleus distance from the dopant for the different values of the magnetic field previously considered. We here represent 500 ions ranging 3.4\AA\, (nearest neighbour distance) to 18.5\AA. As a reminder, in modeling Fig. \ref{fig:mods_fig_fit}, we only use 100 ions (from  3.4\AA\, to 8.6\AA). The red dashed lines at 50mT serves for the approximate spherical model  that will be discussed in section \ref{analytical} where we assume an effective constant branching contrast (50mT, left) and a truncated dipolar splitting decay (50mT, right).}
\label{fig:rho_splittings}
\end{figure}

The superhyperfine induced decay can be interpreted by reconsidering \eqref{Vi} for the different yttriums. We plot the key parameters $\rho_i$, $\Delta_{i}$ and $\Delta^\prime_{i}$ as a function of the distance from the electron spin in Fig.\ref{fig:rho_splittings}.

The curves appear erratic because of the superhyperfine interaction anisotropy. Indeed, for the same distance, certain yttriums have very different angular coordinates, so they may have very different splittings.
Despite the irregular nature of the branching contract (left column of Fig.\ref{fig:rho_splittings}) and the splittings (right column of Fig.\ref{fig:rho_splittings}), the plots can be analyzed as follows. The total magnetic field contains two contributions. First, the dipolar field generated by the \er which globally decreases as $1/r^3$ (neglecting the orientational dependency in a first approach). Second, the constant bias field so the splittings  tend asymptomatically at large distances to the nuclear spin Zeeman splitting (2.1MHz/T) whatever the dopant in the ground or excited state. The branching contrast can only be significant if the dipolar field dominates the bias field \cite{car2017selective, car:tel-02339801} (see also \ref{appendix:general}). In a sense, the bias magnetic field screens the area of influence of electron spin. At a certain distance, the magnetic field becomes larger than the \er dipolar field, so the branching of distant yttriums is very weak.

One could for example define a screening radius within which the \er field dominates the bias field and leads to a large branching contrast. The broadband excitation of the \y ions under the influence of \er (namely with a significant branching contrast) explains the collapse of the echo signal. The number of yttriums increases as the magnetic field is reduced (thus reducing the screening). This will be discussed quantitatively in \ref{analytical}.

At the extreme, the sphere of influence covers the whole space as the magnetic field goes to zero. The number of nuclei diverges with a relatively low average branching value (typically $\sim 0.1$ with large fluctuations). Even though, there is no singularity in the modulation pattern because the splittings decay rapidly by following a $1/r^3$ law (at 0 mT, see Fig.\ref{fig:rho_splittings}).

As mentioned earlier, we have extended the results of our model to 0 mT in Fig.\ref{fig:rho_splittings} with a satisfying agreement with the data set in Fig.\ref{fig:mods_fig_fit}. Nonetheless, this zero-field analysis should be handle with precaution. In our model, we indeed assume that the lowest spin states of $^{4}I_{15/2} $ and $^{4}I_{13/2}$ are selectively addressed. The magnet current is reduced to zero to obtain the 0 mT curve. The spin state transitions are not optically resolved anymore, so our model doesn't strictly apply.
In any case, the expectation values of the \er dipole moment are only well defined (See appendix \ref{appendix:C2} for details) if the magnetic field is present to align the spins. Even when the magnet current is reduced to zero, one cannot exclude the presence of a remanent field because of a magnetization of the sample holder parts or the earth magnetic field. Nevertheless, there is no reason for this remanent field to be aligned with the bias magnetic field used in \ref{sec:model}. So the extension of the model for a given dipole moment orientation at zero-field and the fortunate agreement with the measurement should deserve more investigations.

\section{Approximate spherical model}\label{analytical}

The goal of this section is to move away from the accurate heavy calculations and to give some physical content to the different parameters, primarily $\rho_{i}$, $\Delta_{i}$ and $\Delta^\prime_{i}$ in \eqref{Vi}. There are two features that we would like to put forward by focusing on one set of data in Fig.\ref{fig:mods_fig_fit} at 50mT. First, the rapid initial decay time is a consequence of quasi-continuum of superhyperfine splittings when a large \y ensemble is excited by the echo sequence. Second, at low fields, the decay curves exhibit a series of revivals, that cannot be qualified as a pure oscillation, as observed in Fig.\ref{fig:mods_fig_fit} at 50mT (two revivals between $t_{12}=0$ and $25\mu s$ ) and in the complementary measurements in Fig.\ref{fig:mods_fig_fit_appendix} at 17mT (a single revival in the range $t_{12}\sim 25\mu s$).

Both aspects can be addressed analytically by introducing a continuous distribution of superhyperfine splittings and defining a \er sphere of influence whose radius depends on the magnetic field as we will see now.

As discussed qualitatively in \ref{sec:interpretation}, we observe that the dipolar field globally decreases as $1/r^3$ in both ground and excited states up to a point where it is dominated by the bias field (so the splittings tend to the nuclear Zeeman values). This defines a screening radius for the \er dipolar field. We can then  make a crude assumption. Let's assume that all the nuclei for which the dipolar field dominates the bias magnetic field have a non-zero branching contrast. On the contrary, when the bias field dominates, the branching is zero. This defines a hard sphere of influence of the \er ion. In other words, out of a certain screening radius, \y are assumed completely decoupled from the electron spin.
The \er ground and excited dipolar fields cannot be strictly equal otherwise the branching contrast would be zero \cite{car2017selective, car:tel-02339801} (see also \ref{appendix:general}). In other words, the ground and excited dipolar fields are equal to the lowest order but slightly misaligned to generate a weak branching contrast to the first order as observed in Fig.\ref{fig:rho_splittings} (50mT) with $\rho_i \ll 1$.

In practice, we write the splittings of the \y numbered $i$ at the distance $r$ as $\Delta_{i} \simeq \Delta^\prime_{i} \simeq \bar{\Delta}$ which follows the $1/r^3$ decay law as

\begin{align}
\bar{\Delta}\left( r \right)  =\Delta_0 \frac{r_0^3}{r^3}
\label{r3_law}
\end{align}

where $r_0=3.4$\AA, the nearest neighbour distance and $\Delta_0$ the corresponding splitting.  We then assume the branching contrast to be constant  $\rho_{i} \simeq \bar{\rho}$ up to the screening radius $r_S$ and zero elsewhere for $r_i>r_S$

We will keep $\Delta_0$, $ \bar{\rho}$ and $r_S$ as free parameters for the approximate model. Nevertheless, we expect $\Delta_0$ and $\bar{\rho}$ to be of the order of $\Delta_0 \sim 2\pi \times  600\mathrm{kHz}$ and $\bar{\rho} \sim 0.1 $ as observed  in Fig.\ref{fig:rho_splittings} (50mT). Concerning the screening radius $r_S$, we expect that the dipolar splitting $\bar{\Delta}\left( r_S \right)$ at the distance $r_S$ defined as

\begin{equation}
\bar{\Delta}\left( r_S \right) =\Delta_0 \frac{r_0^3}{r_S^3} = \Delta_S
\label{eq:screening}
\end{equation}
to be of the order of $\displaystyle \frac{{\mu}_{\rm Y}}{h} B $  with ${\mu}_{\rm Y}/h$ the \y nuclear (isotropic) dipole moment (expressed in Hz/T). In other words, $r_S$ corresponds to a compensation between the dipolar field and the bias field $B$.

The superhyperfine modulation envelope can now be evaluated. \eqref{Modulation_formule} can be simplified when the branching contrast is small as

\begin{equation}\left[ \mathcal{V}_\mathrm{tot} \left( t_{12}\right) \right]^2 \simeq \exp \left( - \sum_i \bar{\rho}\left[1-\cos\left( {\Delta_i t_{12}} \right) \right]^2 \right)
\end{equation}

The discrete sum is replaced by a continuous integral in the spherical model:

\begin{align}\label{continuous_integral}
&\left[ \mathcal{V}_\mathrm{tot} \left( t_{12}\right) \right]^2 \simeq \\ \nonumber
&\exp \left( -  \int_{r_0}^{r_S} \bar{\rho}\left[1-\cos\left( \bar{\Delta}\left( r \right) t_{12} \right) \right]^2 4\pi r^2 n_Y \mathrm{d}r \right)
\end{align}
where $n_Y=1.83.10^{22}$ at/cm$^3$ is the yttrium density.

The expected modulation pattern calculated as a continuous integral in \eqref{continuous_integral} is truncated to $r_S$ where the branching contrast is assumed to be zero.

The superhyperfine decay can be rewritten as
\begin{equation}\label{almost_shf_decay}
\exp \left( -\frac{8\pi}{3} n_Y  r_0^3   \bar{\rho} {\Delta_0} t_{12}  \int_{{\Delta_S} t_{12}/2}^{{\Delta_0} t_{12}/2} \frac{\sin^4\phi}{\phi^2}  \mathrm{d}\phi \right)
\end{equation}
using the change of variable $\phi=\displaystyle \left( \Delta_0 t_{12}/2 \right) \frac{r_0^3}{r^3}$ where we have introduced the value of  ${\Delta_S}$ defined in \eqref{eq:screening}

The integral term $\displaystyle \int_{{\Delta_S} t_{12}/2}^{{\Delta_0} t_{12}/2} \frac{\sin^4\phi}{\phi^2}  \mathrm{d}\phi$ oscillates as a function of $t_{12}$ and modulates the exponential decay term given by $\displaystyle  \frac{8\pi}{3} n_Y  r_0^3   \bar{\rho} {\Delta_0} t_{12} $. In a sense, the latter gives a characteristic decay time of $\displaystyle  \frac{3}{8\pi n_Y  r_0^3   \bar{\rho} {\Delta_0} }$.

\begin{figure}[h]
\centering
\includegraphics[width=\columnwidth]{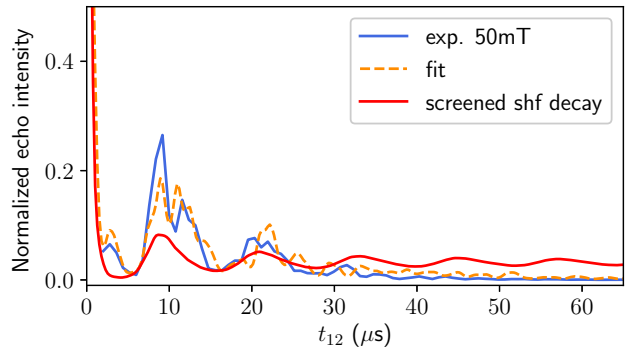} 
\caption{Revivals of the echo decay (blue line) at 50mT (as in Fig.\ref{fig:mods_fig_fit}, the orange dashed line is the theoretical formula \eqref{Modulation_formule}. The qualitative shape is reproduced by introducing a screening of the superhyperfine interaction in \eqref{almost_shf_decay} (screened superhyperfine decay in red line, see text for analysis).}
\label{fig:almost_zerofield_decay}
\end{figure}

This expression can phenomenologically reproduce the oscillating decay curve  at 50mT. To do so, we leave ${\Delta_0}$, ${\Delta_S}$ and $\bar{\rho}$ as free parameters and fit the experimental data (blue line in Fig.\ref{fig:mods_fig_fit}, 50mT). The agreement in Fig.\ref{fig:almost_zerofield_decay} is qualitatively satisfying.

The best fit gives reasonable values of $\Delta_0=2\pi \times  635 \mathrm{kHz} $ and 
$\Delta_S =\displaystyle 0.8 \frac{{\mu}_{\rm Y}}{h} B $ for the screening splitting, and $\bar{\rho} = 0.11$ for the contrast as expected from the analysis of Fig.\ref{fig:rho_splittings} (50mT).

More importantly, the value $\Delta_S =\displaystyle 0.8 \frac{{\mu}_{\rm Y}}{h} B $ corresponds well to our expectation. When the splitting are larger than $\Delta_S$, the dipolar field coarsely dominates the nuclear Zeeman term. This is the way we have defined the screening sphere of the \er influence. From $\Delta_S$, we can extract the value of the screening radius $r_S=6.7$ \AA\, using \eqref{eq:screening}. This should be compared to $r_0=3.4$\AA, the nearest neighbor distance. 

\eqref{eq:screening} is plotted with the fitted parameters in  Fig.\ref{fig:rho_splittings} (50mT, right column, dashed red line) for comparison with the accurate model of section \ref{sec:model} between $r_0=3.4$ \AA\, and $r_S=6.7$ \AA. The fitted value of the branching $\bar{\rho} = 0.11$ is also represented  in  Fig.\ref{fig:rho_splittings} (50mT, left column, red dashed line). Theses effective values of the splittings and the branching contrast allow to reproduce satisfyingly our case of study at 50mT.

As the magnetic field is increased, the radius decreases thus reducing the number of interacting nuclei. This justifies the need to apply a minimal magnetic field. Indeed, the superhyperfine induced decay regime can be eliminated by increasing the field to a point where the screening radius is comparable to $r_0$ so a very limited number of \y are still interacting. Additionally, when a few ions are in the \er area of influence, the anisotropy of the electron spin can be used to turn on or off the interaction with isolated nuclei \cite{car2017selective}.

\section{Conclusion}

We have used the photon-echo technique to investigate the decoherence of \eryso at low field. The collapse of the coherence time is not due to a modification of the electron or nuclear spins dynamics that is essentially unaffected at low field. Instead, the decay curves are accurately explained by the superhyperfine modulations that involve an increasing number of nuclei as the field is reduced. The low-field decay can be reproduced analytically by considering a spherical model (angular averaging) and introducing a cut-off of the \er dipolar field, thus defining a screening radius of the electron spin influence.

The term decoherence for the superhyperfine induced collapse is actually questionable. During the {\it true}  coherence time, $T_2=58 \mu$s in our case, the evolution of the spin ensemble (\er and a large collection of nuclei) is indeed unitary and potentially reversible. This is a striking feature of the mesoscopic ensemble evolution (cluster of \y around \er). One may wonder if the rapid superhyperfine decay can be cancelled thus exploiting the reversibility of the process. There is no obvious solution except extending the pulse duration or reducing the power to perform a spectral selection \cite{braunschweiler1985selective, barkhuijsen1985partial, astashkin1987dependence}. This is an interesting approach to gain understanding on the system but this constrains the {\it apparent} decay to the experimental parameters (pulse duration for example). 

This doesn't necessarily mean that the superhyperfine induced collapse should be considered as a hard limit. One could on the contrary consider the repetition of short pulses to compensate for the dephasing induce by the inhomogeneous superhyperfine couplings. Despite a clear analogy with the dynamical decoupling technique, the term dynamical is not appropriate because the {\it apparent} decay is not driven by the dynamical fluctuations of the environment. Additionally, the transposition of this RF technique to the optical domain is not direct because repeated coherence refocusing would trigger the emission of multiple photon-echos. Still, our analysis shows that the application of sub-$\mu$s pulses would compensate for the superhyperfine collapse. They are not technically accessible in our case because a  large peak power is needed to maintain a significant pulse area. This nevertheless draws a stimulating perspective to compensate for the superhypefine coupling to a large nuclear spin ensemble at low field.

\section*{Acknowledgements}
We thank O. Arcizet for lending the extended cavity diode laser used for the measurements and the technical help of the team QuantECA on cryogenics.

We have received funding from the Investissements d'Avenir du LabEx PALM ExciMol, ATERSIIQ and OptoRF-Er (ANR-10-LABX-0039-PALM). This work was supported by the ANR MIRESPIN project, grant ANR-19-CE47-0011 of the French Agence Nationale de la Recherche.

\bibliography{mods_bib}{}

\clearpage
\appendix
\section{Complementary measurements}\label{sec:complementary}

\begin{figure}[h]
\centering
\includegraphics[width=\columnwidth]{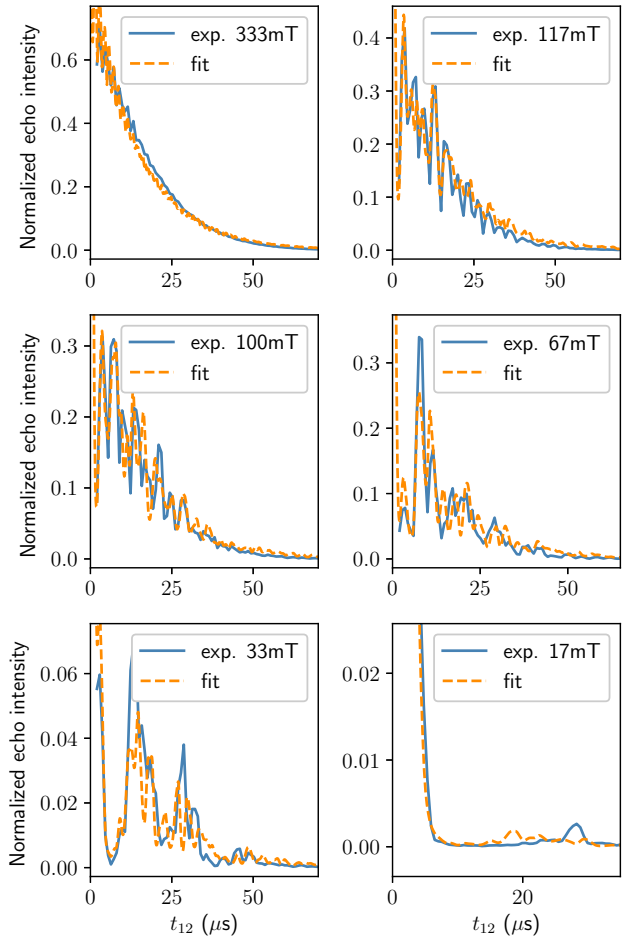} 
\caption{Complementary experimental measurements echo decay curves (blue line) with different magnetic field values than Fig. \ref{fig:mods_fig_exp}. As in Fig. \ref{fig:mods_fig_fit}, we have superimposed the theoretical formula \eqref{Modulation_formule} (orange dashed line). The fit is normalized to one at $t_{12}=0$. We have adjusted the vertical scale to draw the attention on the experimental data.}
\label{fig:mods_fig_fit_appendix}
\end{figure}

We complement the experimental measurements of Fig. \ref{fig:mods_fig_exp} with different magnetic field values. The fitting procedure explained in section \ref{sec:model} , so Fig.\ref{fig:mods_fig_fit_appendix} is analogous to Fig. \ref{fig:mods_fig_fit}.

All the curves (Fig. \ref{fig:mods_fig_fit} and Fig.\ref{fig:mods_fig_fit_appendix}) are fit with a unique coherence time $T_2=58 \mu$s.

\section{Calculation of the modulation parameters}\label{appendix:C2}
We first remind the general reasoning to calculate the modulation parameters of \eqref{Vi} in \ref{appendix:general}. The calculation has been already detailed in \cite{car2017selective}. We also treat the inequivalent site orientations, sometime called magnetic subsites, present in the \yso matrix. The  fit agreement in Fig.\ref{fig:mods_fig_fit} is obtained when the magnetic doesn't exactly lies in the $(D_1,D_2)$ plane but slightly offset. In that case, the two possible site orientations (called orientations I and II in \cite{sun_magnetic_2008} and related by a C$_2$ rotation about $b$) are not equivalent and have to be treated independently (see \ref{appendix:subsites}).

\subsection{General approach}\label{appendix:general}

For a given \y spin whose location is $\vec{r}$ referenced from the dopant, the effect of the \er electron spin can be treated in a perturbative manner by introducing the Hamiltonian  \begin{equation}\displaystyle H_{g,e}^\prime  = - \vec{\mu}_{\rm Y} \cdot \vec{B}_{g,e} \left( \vec{r} \right)  \end{equation} depending if the \er is in the ground or the optical excited state (written with subscripts $g$ or $e$) respectively \cite{car2017selective}. $\vec{\mu}_{\rm Y}$ is the \y nuclear (isotropic) dipole moment.
The field 
\begin{equation} \vec{B}_{g,e} \left( \vec{r} \right) = \vec{B} + \vec{B}^{\rm Er}_{g,e} \left( \vec{r} \right) \end{equation} is the total magnetic field seen by the \y ion including the bias field $\vec{B}$ and the dipolar field generated by the \er spin at the location $\vec{r}$ of the nuclear spin. The latter takes the general form
\begin{equation}\label{B_dipole}
\vec{B}^{\rm Er}_{g,e} = - \frac{\mu_0}{4\pi} \left[ \frac{\langle \vec{\mu}^{\rm Er}_{g,e} \rangle }{r^3}-3\frac{\left(\langle \vec{\mu}^{\rm Er}_{g,e} \rangle  \cdot\vec{r}\right)\cdot\vec{r}}{r^5}\right]
\end{equation}
where $\langle \vec{\mu}^{\rm Er}_{g,e} \rangle$ is the \er electron spin dipole moment (expectation value), written $\vec{\mu}=\langle \vec{\mu}^{\rm Er}_{g} \rangle$ (ground) and $\vec{\mu^\prime}=\langle \vec{\mu}^{\rm Er}_{e} \rangle$ (excited) in the text.

This is sufficient to extract the parameters of interest in \eqref{Vi}, namely $\Delta$ and $\Delta^\prime$ the superhyperfine splittings when the erbium is in the ground or excited state as the eigenvalues of $H_{g}^\prime$ and $H_{e}^\prime$ respectively. Finally, the branching contrast $\rho$ has the simple geometrical interpretation as $\rho = \sin^2 \left( \theta \right)$ where $\theta = \left( \vec{B}_{g},\vec{B}_{e} \right)$, the angle between the total magnetic fields.

The dipole moments $\vec{\mu}$ and $\vec{\mu^\prime}$ are not aligned with the magnetic field $ \vec{B}$. This is a consequence of the local site anisotropy as revealed by the anisotropy of the $g$-tensor tabulated in \cite[Table.III]{sun_magnetic_2008}.

\subsection{Considering the two possible magnetic orientations}\label{appendix:subsites}

If the magnetic field $ \vec{B}$ is contained in the $(D_1,D_2)$ plane or parallel to $b$, then there is no need to bother with the two different site orientations because their are equivalent. In short, the magnetic subsites have the same dipoles moments $\vec{\mu}$ and $\vec{\mu^\prime}$.

Otherwise, the orientations I and II have to be treated independently. There is no special difficulty. The corresponding $g$-tensors are related by a C$_2$ rotation about $b$ as tabulated in \cite[Table.III]{sun_magnetic_2008}. The values of $\vec{\mu}$ now written $\vec{\mu}_\mathrm{I}$ and  $\vec{\mu}_\mathrm{II}$ are calculated independently (same for $\vec{\mu^\prime}$). We do not obtain the same values for $\vec{B}^{\rm Er}_{g}$ in \eqref{B_dipole}, obviously because the $\vec{\mu}_\mathrm{I}$ and  $\vec{\mu}_\mathrm{II}$ are different (same for  $\vec{B}^{\rm Er}_{e}$ with $\vec{\mu^\prime}$) but also because the relative positions of the surrounding \y are different: the C$_2$ rotation has to be applied to $\vec{r}$ for each yttrium. We give the explicit positions of 100 ions in \ref{appendix:positions} that may serve to the reader for further analysis.

We finally calculate the parameters $\rho_{i}$, $\Delta_{i}$ and $\Delta^\prime_{i}$ in \eqref{Vi} for an ensemble of 100 ions  (labeled $i$) per orientation and obtain the modulations $\mathcal{V}_\mathrm{tot}^\mathrm{I}$ and $\mathcal{V}_\mathrm{tot}^\mathrm{II}$ for each orientation. We sum the two contributions in a coherent manner to evaluate the final ensemble modulation
\begin{equation}
\mathcal{V}_\mathrm{tot}=\frac{\mathcal{V}_\mathrm{tot}^\mathrm{I}+\mathcal{V}_\mathrm{tot}^\mathrm{II}}{2}
\end{equation}

\section{Positions of the \y ions}\label{appendix:positions}
We here give the \y positions used to calculate the superfhyperfine splittings in orientation I. We write the first 20 ions (sorted by distance from the \er center) in table \ref{tab:positions}. The coordinates are given in the frame $D_1$,$D_2$ and $b$. The complete set of 100 ions used for the calculation is given as supplementary materials in the form of a text file.

\begin{table}[h]
\centering
\caption{\y positions for orientation I (\AA)}
\begin{tabular}{ | c | c | c | c | c| }
\hline
\y number &  distance $r$ & $D_1$ coord. & $D_2$ coord. & $b$ coord.  \\
  \hline \hline  
1 & 3.40 & -0.66 & 3.23 & -0.81 \\ \hline
2 & 3.46 & -3.45 & 0.28 & 0.00 \\ \hline
3 & 3.51 & -1.66 & -1.88 & 2.45 \\ \hline
4 & 3.62 & 2.27 & -2.24 & -1.72 \\ \hline
5 & 3.72 & -1.79 & 2.15 & 2.45 \\ \hline
6 & 4.15 & -2.79 & -2.95 & -0.81 \\ \hline
7 & 4.70 & 3.93 & -0.37 & 2.55 \\ \hline
8 & 4.95 & -1.66 & -1.88 & -4.27 \\ \hline
9 & 5.10 & -1.79 & 2.15 & -4.27 \\ \hline
10 & 5.19 & 5.06 & 0.71 & -0.91 \\ \hline
11 & 5.46 & -1.01 & -5.11 & 1.64 \\ \hline
12 & 5.46 & 1.01 & 5.11 & 1.64 \\ \hline
13 & 5.50 & 3.27 & 2.86 & -3.36 \\ \hline
14 & 5.50 & 3.27 & 2.86 & 3.36 \\ \hline
15 & 5.74 & 3.93 & -0.37 & -4.17 \\ \hline
16 & 5.93 & 2.27 & -2.24 & 5.00 \\ \hline
17 & 6.14 & -2.44 & 5.38 & 1.64 \\ \hline
18 & 6.27 & 2.92 & -5.47 & -0.91 \\ \hline
19 & 6.48 & -5.71 & 2.52 & -1.72 \\ \hline
20 & 6.48 & 5.71 & -2.52 & -1.72 \\ \hline
\end{tabular}
  \label{tab:positions}
\end{table}

We also give the first five \y ions  in orientation II in table \ref{tab:positions2}. They are simply deduced from the positions for the other site by a  C$_2$ rotation about $b$.

\begin{table}[h]
\centering
\caption{\y positions for orientation II (\AA)}
\begin{tabular}{ | c | c | c | c | c| }
\hline
\y number &  distance $r$ & $D_1$ coord. & $D_2$ coord. & $b$ coord.  \\
  \hline \hline  
1 & 3.40 & 0.66 & -3.23 & -0.81 \\ \hline
2 & 3.46 & 3.45 & -0.28 & 0.00 \\ \hline
3 & 3.51 & 1.66 & 1.88 & 2.45 \\ \hline
4 & 3.62 & -2.27 & 2.24 & -1.72 \\ \hline
5 & 3.72 & 1.79 & -2.15 & 2.45 \\ \hline
$\cdots$ &$\cdots$ &$\cdots$ &$\cdots$ & $\cdots$  \\ \hline
\end{tabular}
  \label{tab:positions2}
\end{table}

\newpage
\section{Supplementary material: Complete set of 100 ions used for the calculation}
\input{table_positions_I_raw_arxiv.txt}

\end{document}

%% file: table_positions_I_raw_arxiv.txt
distance r ; D1 coord. ; D2 coord. ; b coord.\\
3.39533 ; 0.65503 ; -3.23077 ; -0.81324 \\
3.45926 ; 3.44812 ; -0.27745 ; 0.00000 \\
3.50687 ; 1.66157 ; 1.87598 ; 2.45316 \\
3.62270 ; -2.26581 ; 2.24269 ; -1.72058 \\
3.72116 ; 1.78655 ; -2.15343 ; 2.45316 \\
4.14545 ; -2.79308 ; -2.95332 ; -0.81324 \\
4.69545 ; 3.92737 ; -0.36671 ; 2.54726 \\
4.94920 ; -1.66157 ; -1.87598 ; -4.26783 \\
5.10328 ; -1.78655 ; 2.15343 ; -4.26783 \\
5.18851 ; 5.05889 ; 0.71063 ; -0.90734 \\
5.45723 ; -1.00653 ; -5.10675 ; 1.63992 \\
5.45723 ; 1.00653 ; 5.10675 ; 1.63992 \\
5.49582 ; 3.27234 ; 2.86406 ; -3.36050 \\
5.49582 ; 3.27234 ; 2.86406 ; 3.36050 \\
5.74272 ; 3.92737 ; -0.36671 ; -4.17374 \\
5.93024 ; 2.26581 ; -2.24269 ; 5.00042 \\
6.13517 ; -2.44158 ; 5.38420 ; 1.63992 \\
6.27004 ; 2.92084 ; -5.47346 ; -0.90734 \\
6.47769 ; -5.71392 ; 2.52014 ; -1.72058 \\
6.47769 ; 5.71392 ; -2.52014 ; -1.72058 \\
6.49893 ; 4.05236 ; -4.39612 ; 2.54726 \\
6.72100 ; 0.00000 ; 0.00000 ; -6.72100 \\
6.72100 ; 0.00000 ; 0.00000 ; 6.72100 \\
6.76525 ; -0.65503 ; 3.23077 ; 5.90776 \\
6.77166 ; -4.45465 ; -4.82931 ; 1.63992 \\
6.90179 ; 4.93391 ; 4.74004 ; -0.90734 \\
7.17113 ; -2.79308 ; -2.95332 ; 5.90776 \\
7.27388 ; -1.00653 ; -5.10675 ; -5.08108 \\
7.27388 ; 1.00653 ; 5.10675 ; -5.08108 \\
7.29161 ; 4.05236 ; -4.39612 ; -4.17374 \\
7.34711 ; -0.78002 ; 7.26018 ; -0.81324 \\
7.51922 ; -2.66810 ; -6.98274 ; -0.81324 \\
7.55899 ; -3.44812 ; 0.27745 ; -6.72100 \\
7.55899 ; -3.44812 ; 0.27745 ; 6.72100 \\
7.73925 ; 5.05889 ; 0.71063 ; 5.81367 \\
7.79540 ; -2.44158 ; 5.38420 ; -5.08108 \\
7.82952 ; -7.37549 ; 0.64416 ; 2.54726 \\
7.94658 ; -6.72046 ; -2.58661 ; -3.36050 \\
7.94658 ; -6.72046 ; -2.58661 ; 3.36050 \\
7.94658 ; 6.72046 ; 2.58661 ; -3.36050 \\
7.94658 ; 6.72046 ; 2.58661 ; 3.36050 \\
8.00027 ; -5.71392 ; 2.52014 ; 5.00042 \\
8.00027 ; 5.71392 ; -2.52014 ; 5.00042 \\
8.17882 ; 1.25927 ; -7.34945 ; -3.36050 \\
8.17882 ; 1.25927 ; -7.34945 ; 3.36050 \\
8.22385 ; -0.65503 ; 3.23077 ; -7.53424 \\
8.30563 ; -4.45465 ; -4.82931 ; -5.08108 \\
8.49899 ; -7.37549 ; 0.64416 ; -4.17374 \\
8.50228 ; 2.92084 ; -5.47346 ; 5.81367 \\
8.56085 ; -2.79308 ; -2.95332 ; -7.53424 \\
8.56622 ; -8.50701 ; -0.43318 ; -0.90734 \\
8.62901 ; -6.36896 ; 5.75091 ; -0.90734 \\
8.69803 ; 0.35150 ; 8.33752 ; 2.45316 \\
8.78179 ; 6.06542 ; 5.81738 ; 2.54726 \\
8.97832 ; 4.93391 ; 4.74004 ; 5.81367 \\
9.02351 ; 2.26581 ; -2.24269 ; -8.44158 \\
9.14115 ; 8.98626 ; 0.34392 ; 1.63992 \\
9.18090 ; 5.05889 ; 0.71063 ; -7.62833 \\
9.19717 ; -7.50047 ; 4.67357 ; 2.54726 \\
9.20885 ; 4.27887 ; 7.97081 ; -1.72058 \\
9.24229 ; -3.79962 ; -8.06007 ; 2.45316 \\
9.29304 ; 7.97973 ; -4.76283 ; 0.00000 \\
9.37295 ; 0.35150 ; 8.33752 ; -4.26783 \\
9.38357 ; 6.06542 ; 5.81738 ; -4.17374 \\
9.39257 ; -0.78002 ; 7.26018 ; 5.90776 \\
9.51028 ; -1.66157 ; -1.87598 ; 9.17417 \\
9.52780 ; -2.66810 ; -6.98274 ; 5.90776 \\
9.53920 ; -8.38202 ; -4.46260 ; -0.90734 \\
9.57194 ; -4.70739 ; 7.62689 ; 3.36050 \\
9.57194 ; 4.70739 ; -7.62689 ; -3.36050 \\
9.57194 ; 4.70739 ; -7.62689 ; 3.36050 \\
9.57194 ; -4.70739 ; 7.62689 ; -3.36050 \\
9.59137 ; -1.78655 ; 2.15343 ; 9.17417 \\
9.73293 ; -9.16204 ; 2.79759 ; -1.72058 \\
9.77341 ; -7.50047 ; 4.67357 ; -4.17374 \\
9.83268 ; 2.92084 ; -5.47346 ; -7.62833 \\
9.84871 ; -1.00653 ; -5.10675 ; 8.36092 \\
9.84871 ; 1.00653 ; 5.10675 ; 8.36092 \\
9.88010 ; -3.79962 ; -8.06007 ; -4.26783 \\ 
10.02023 ; 3.04582 ; -9.50288 ; -0.90734 \\
10.07270 ; 3.92737 ; -0.36671 ; 9.26826 \\
10.23992 ; -2.44158 ; 5.38420 ; 8.36092 \\
10.24709 ; 4.93391 ; 4.74004 ; -7.62833 \\
10.31288 ; -8.50701 ; -0.43318 ; 5.81367 \\
10.32901 ; 8.98626 ; 0.34392 ; -5.08108 \\
10.33667 ; 4.27887 ; 7.97081 ; 5.00042 \\
10.35889 ; 9.64130 ; -2.88685 ; 2.45316 \\
10.36510 ; -6.36896 ; 5.75091 ; 5.81367 \\
10.41000 ; -2.01307 ; -10.21350 ; 0.00000 \\
10.41000 ; 2.01307 ; 10.21350 ; 0.00000 \\
10.49207 ; -0.78002 ; 7.26018 ; -7.53424 \\
10.50049 ; -5.71392 ; 2.52014 ; -8.44158 \\
10.50049 ; 5.71392 ; -2.52014 ; -8.44158 \\
10.58865 ; -1.43505 ; 10.49095 ; 0.00000 \\
10.61330 ; -2.66810 ; -6.98274 ; -7.53424 \\
10.63349 ; -4.45465 ; -4.82931 ; 8.36092 \\
10.80619 ; -9.16204 ; 2.79759 ; 5.00042 \\
10.90714 ; 10.64783 ; 2.21990 ; -0.81324 \\
10.93174 ; 9.64130 ; -2.88685 ; -4.26783 \\
10.95396 ; 10.77281 ; -1.80951 ; -0.81324 \\